\documentclass[11pt]{article}
\usepackage{verbatim,amsmath,amssymb}
\usepackage{epsfig,float,color}
\usepackage{epstopdf}
\usepackage{geometry}
\usepackage{setspace}
\usepackage{wrapfig}
\usepackage{hyperref}
\usepackage[utf8]{inputenc}
\usepackage{graphicx}
\usepackage{flushend}
\usepackage[linesnumbered,ruled]{algorithm2e}

\usepackage[font={footnotesize}]{caption}
\usepackage[font={footnotesize}]{subcaption}
\usepackage{footnote}
\usepackage{multicol}
\usepackage{multirow}
\usepackage{enumitem}   
\usepackage{soul}
\usepackage{framed}
\usepackage[titletoc,title]{appendix}
\usepackage{bm}
\usepackage{siunitx}
\usepackage{cite}
\usepackage[makeroom]{cancel} 
\usepackage[authoryear]{natbib}
\usepackage{hyperref}
\definecolor{darkblue}{rgb}{0,0,1}
\hypersetup{pdftex=true, colorlinks=true, breaklinks=true, linkcolor=magenta, menucolor=magenta, citecolor=magenta, urlcolor=magenta}

\renewcommand{\bar}[1]{\tilde{#1}}

\definecolor{darkblue}{rgb}{0,0,1}
\hypersetup{pdftex=true, colorlinks=true, breaklinks=true, linkcolor=darkblue, menucolor=darkblue, pagecolor=darkblue, citecolor=darkblue, urlcolor=darkblue}

\usepackage[dvipsnames]{xcolor}
\usepackage{tikz,pgfplots}
\usetikzlibrary{calc, pgfplots.groupplots, backgrounds}
\usepgfplotslibrary{colorbrewer}
\usepackage{standalone}

\begin{document}
	
	\begin{center}
		\Large{\bf{On topology optimization of large deformation contact-aided shape morphing compliant mechanisms}}\\
		
	\end{center}
	
	\begin{center}
			\large{Prabhat Kumar\,$^{\star,\,\ddagger,}$\footnote[1]{Corresponding author, email: \url{prabhatkumar.rns@gmail.com}}, Roger A. Sauer$\,^{\curlyvee,\,\ast,\,\dagger}\,$ and Anupam Saxena\,$^{\dagger}$} \\
		\vspace{4mm}
		
		\small{\textit{$\star$Department of Mechanical Engineering, Solid Mechanics, Technical University of Denmark,
				2800 Kgs. Lyngby, Denmark}}\\
			\vspace{2mm}	
		\small{\textit{$\ddagger$ Faculty of Civil and Environmental Engineering, Technion-Israel Institute of
				Technology, Haifa, Israel}}\\
				\vspace{2mm}
		\small{\textit{$\curlyvee$Graduate School, AICES, RWTH
				Aachen University, Templergraben 55, 52056 Aachen, Germany}}\\
				\vspace{2mm}
		\small{\textit{$\ast$Faculty of Civil and Environmental Engineering, Gda\'{n}sk University of Technology, ul. Narutowicza 11/12, 80-233 Gda\'{n}sk, Poland}}\\
			\vspace{2mm}
		\small{\textit{$\dagger$Department of Mechanical Engineering, Indian Institute of Technology Kanpur, UP 208016, India}}\\
		
		\vspace{4mm}
Published\footnote{This pdf is the personal version of an article whose final publication is available at \href{https://www.sciencedirect.com/science/article/pii/S0094114X20303529}{Mechanism and Machine Theory}}\,\,\,in \textit{Mechanism and Machine Theory}, 
\href{https://doi.org/10.1016/j.mechmachtheory.2020.104135}{DOI:10.1016/j.mechmachtheory.2020.104135} \\
Submitted on 08~June 2020, Revised on 10~August 2020, Accepted on 06~October 2020
		
	\end{center}
	
	\vspace{3mm}
	\rule{\linewidth}{.15mm}
	{\bf Abstract:}
	A topology optimization approach for designing large deformation contact-aided shape morphing compliant mechanisms 
	is presented. Such mechanisms can be used in varying operating conditions.  Design domains are described by regular hexagonal elements. Negative circular masks are employed to perform dual task, i.e., to decide material states of each element and also, to generate rigid contact surfaces. Each mask is characterized by five design variables, which are mutated by a zero-order based hill-climbing optimizer. Geometric and material nonlinearities  are considered. Continuity in normals to boundaries of the candidate designs is ensured using a boundary resolution and smoothing scheme. Nonlinear mechanical equilibrium equations are solved using the Newton-Raphson method. An updated Lagrange approach in association with segment-to-segment contact method is employed for the contact formulation. Both mutual and self contact modes are permitted. Efficacy of the approach is demonstrated by designing four contact-aided shape morphing compliant mechanisms for different desired curves. Performance of the deformed profiles is verified using a commercial software. The effect of frictional contact surface on the actual profile is also studied. \\
	
	{\textbf {Keywords:} Shape morphing compliant mechanisms; Topology optimization; Boundary resolution and smoothing; Fourier shape descriptors; Self and mutual contact; Nonlinear finite element analysis}

	\vspace{-4mm}
	\rule{\linewidth}{.15mm}
	
\section{Introduction}\label{Sec:Introduction}
A compliant mechanism (CM), monolithic design, performs its task by deriving motions from elastic deformation of its constituting flexible members. Such mechanisms have many advantages over their traditional linkage-based mechanisms. When  mechanisms also exploit available contact constraints to achieve their objective then those are termed contact-aided compliant mechanisms \citep{mankame2004topology,mankame2007synthesis}. Contact-aided compliant mechanisms (CCMs) can experience either self or mutual (external) or a combination of both contact modes \citep{kumar2019computational}. The former contact occurs when a CM interacts with itself, whereas in the later contact mode, the continuum comes in contact with external (rigid/soft) body. For mutual contact, one can either define external contact surfaces \textit{a priori} \citep{mankame2007synthesis} or generate them systematically \citep{kumar2016synthesis}. However, in case of self contact, one needs to find contact pairs systematically as the members of a candidate design deform and come in contact. A method for contact pairs detection for both contact modes can be found in \cite{kumar2019computational}. One can design CMs and CCMs for a wide range of applications \citep{saxena2001topology,cannon2005compliant,mehta2009stress,reddy2012systematic,tummala2013design,tummala2014design,kumar2017_diss,  kumar2019compliant,kumar2020topology1}.

There exist various design approaches for CMs, which can be broadly classified into: (i) Pseudo Rigid Body Model based approaches \citep{midha1994method,howell2001compliant} and (ii) methods based on topology optimization \citep{ananthasuresh1994strategies,frecker1997topological,sigmund1997design,saxena2000optimal}. Readers may refer to a  review article \citep{zhu2020design} on the approaches for designing CMs using topology optimization. The former approaches employ concepts of kinematics wherein CMs are designed from their initially known rigid-linkage mechanisms. On the other hand, topology optimization based approaches find the optimum material layout of a given design domain with know boundary conditions by extremizing a formulated and/or given objective under a set of known constraints. Generally, a CM should be designed to provide adequate flexibility and also, should sustain under external actuation. One can achieve the later requirement using constraints on either strain energy, or input displacements or maximum stress, etc,  whereas output deformation can be employed to indicate the first measure.  \citet{ananthasuresh1994strategies} formulated a weighted objective using strain-energy and output deformation and extremized that to synthesize CMs. \cite{frecker1997topological} maximized  the ratio of output deformation and strain energy. \cite{saxena2000optimal} generalized the multi-criteria objective. \cite{sigmund1997design} optimized the objective stemming from the mechanical advantage with constraints on volume and input displacements. \cite{saxena2001topology} and \cite{pedersen2001topology} synthesized path generating CMs by extremizing an objective based on a least-square error.  To avoid timing constraints arising naturally in least square based objectives, \cite{ullah1997optimal} employed an objective derived using Fourier Shape Descriptors \citep{zahn1972fourier}.  \cite{rai2007synthesis} used a Fourier Shape Descriptors based objective with curved beam and rigid truss elements to design fully and compliant partially path-generating mechanisms. 

\begin{figure}[h!]
	\centering
	\includegraphics[scale =1]{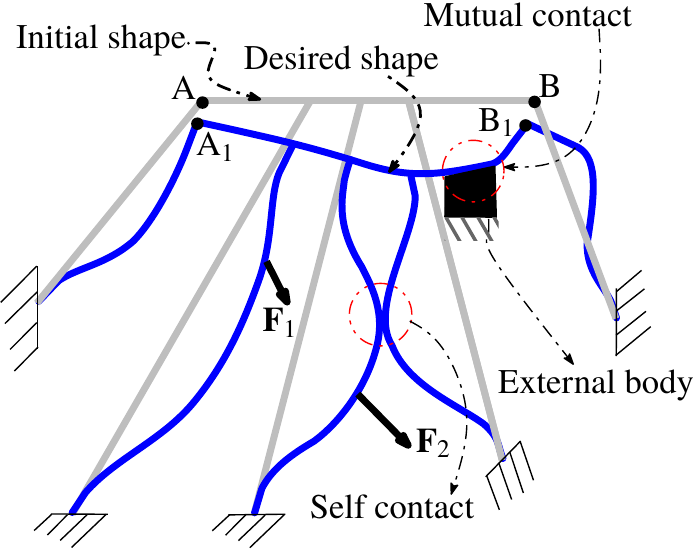}
	\caption{ A schematic diagram for illustrating a contact-aided shape morphing  compliant mechanism. The initial and desired shapes of member AB are shown when the CM is subjected to forces $\mathbf{F}_1\,\text{and}\,\mathbf{F}_2$. The CM experiences self and mutual contact while achieving its desired profile A$_1$B$_1$.}
	\label{fig:schemtic}
\end{figure}

The geometrical shapes of the members of a mechanical design determine its performance. Typically, these shapes are fixed, however permitting shape changes in the design can  enhance efficiency and/or flexibility, e.g., aircraft wings, antenna reflector \citep{saggere1999static}. A shape morphing compliant mechanism (SMCM) attains desired shapes in predefined member(s) in response to external stimuli to further increase its performance. A SMCM can be viewed as a CM having multi-output ports interrelated to each other along \textit{a priori} defined flexible branches. Most of the aforementioned work primarily focused on synthesizing CMs to achieve output at a specific/single location.  \cite{larsen1997design} and  \cite{frecker1999topology} were the first to design CMs with multiple output ports.  The authors in \citep{larsen1997design} minimized the error objective stemming from the prescribed and actual geometrical and mechanical advantages, whereas  the latter ones minimized the modified multi-criteria objective \citep{frecker1997topological}. \cite{saxena2005topology} used a genetic algorithm to design such mechanisms with multi-materials.

SMCMs have various applications wherein the mechanisms have to undergo different operating conditions or experience different external loadings/disturbances, e.g., aircraft wings, antenna reflectors. In addition, such morphing characteristics can be exploited efficiently in association with contact constraints \citep{ramrakhyani2005aircraft,wissa2012passively,tummala2014design}, i.e., contact constraints can further increase the range of application of such mechanisms. \cite{saggere1999static,lu2003design} proposed  synthesis approaches for such mechanisms wherein they employed beam elements to represent the design domain. \cite{mehta2008contact} presented morphing aircraft skin using structures consisting of contact-aided compliant mechanisms. A CCM helps alleviating stresses and achieving high stiffness in the direction perpendicular to the plane of deformation. \cite{ramrakhyani2005aircraft} realized morphing aircraft structure using tendon-actuated compliant cellular trusses. \cite{wissa2012passively} designed and tested passively morphing ornithopter wings which were modeled using compliant splines. The contact analyses in \cite{mehta2008contact,ramrakhyani2005aircraft,wissa2012passively} were performed using  commercial software. A typical shape morphing compliant design undergoes large deformation to achieve its desired profile. In addition, some  members of the SMCM may interact internally (self contact) and also, with external rigid bodies (mutual contact) while deforming (Fig.~\ref{fig:schemtic}). Contact may or may not be essential to large deformation SMCMs. However, having contact constraints included in the approach rather makes the design method more generic and suitable for a set of different applications including or excluding contact. In case contact is not desired, one may have to find and penalize the candidate designs whose constituent members intersect. As a consequence, many potent designs may get ignored, or a desired design may not be obtained. Contact analysis is mandatory in case of (contact-aided) SMCMs wherein a subregion must come into contact with another for the latter to produce the desired shape, e.g., Example~3 and Example~4 (Sec.~\ref{Sec:NumericalExamplesdiscusssion}). The aim is to  present a topology optimization approach to design large deformation SMCMs experiencing self and/or mutual contact using continuum optimization. Those mechanisms are termed  contact-aided shape morphing compliant mechanisms (CSMCMs) herein.

The remainder of the paper is organized as follows. Section~\ref{Sec:Methodology} describes the overall methodology for the presented approach. The problem formulation is reported in Section~\ref{Sec:Problemformulation} wherein boundary smoothing, contact finite element, objective formulation and optimization algorithm are presented. Section~\ref{Sec:NumericalExamplesdiscusssion} presents four contact-aided shape morphing mechanisms, comparison with ABAQUS analyses, performance of the optimized designs with different friction coefficients and pertinent discussions. Lastly, in Section~\ref{Sec:Closure}, conclusions are mentioned.

\section{Methodology}\label{Sec:Methodology}
\begin{figure}[h!]
	\centering
	\includegraphics[scale =1]{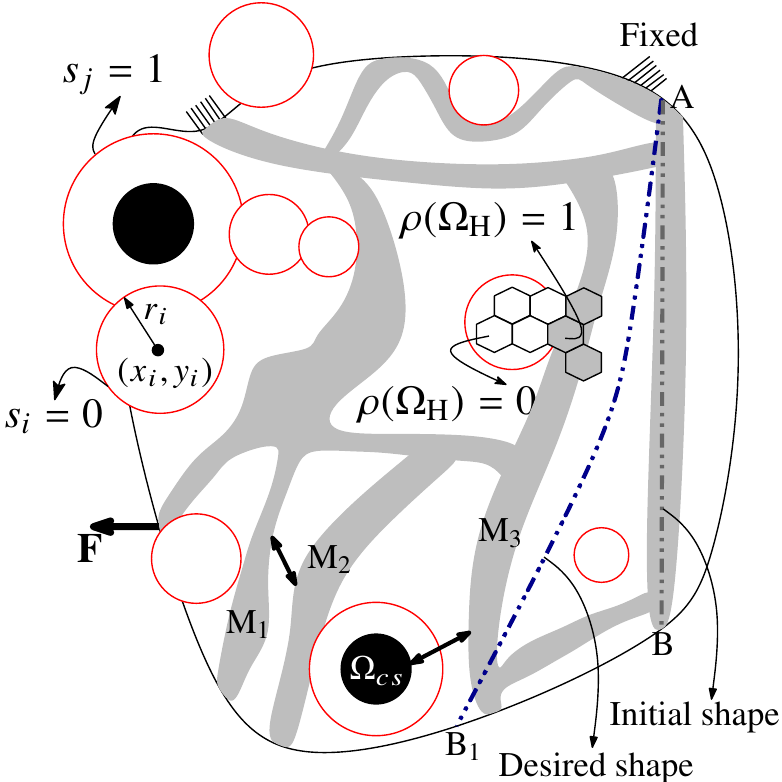}
	\caption{Design procedure for  contact-aided shape morphing compliant mechanisms (CSMCMs). The design domain is discretized by hexagonal elements $\Omega_\mathrm{H}$. Negative circular masks $\Omega_\mathrm{M}$ (red circles) are used to remove material and also, to generate contact surfaces. Five parameters ($x_i,\,y_i,\,r_i,\,s_i,\,f_i $) define each mask. $s_{i}=1$ represents a contact surface (circular solids in black) within the $i^\mathrm{th}$ mask while $s_{i}=0$ implies no contact surface. $\rho(\Omega_\mathrm{H})=0$  implies a void element while $\rho(\Omega_\mathrm{H})=1$ indicates a filled element. Contact surfaces ($\Omega_{cs}$) interact (shown with double head arrows) with the mechanism (mutual contact), e.g., an interaction between member M$_3$ and $\Omega_{cs}$. In addition, the mechanism interacts with itself (self contact), e.g., contact between members M$_1$ and M$_2$. The desired final configuration of the link AB is shown. Fixed boundary(ies) of the domain, input force(s) and output path are also shown.}
	\label{fig:schematicformulation}
\end{figure}
\begin{figure}[h!]
	\begin{subfigure}{0.45\textwidth}
		\centering
		\includegraphics[scale=0.5]{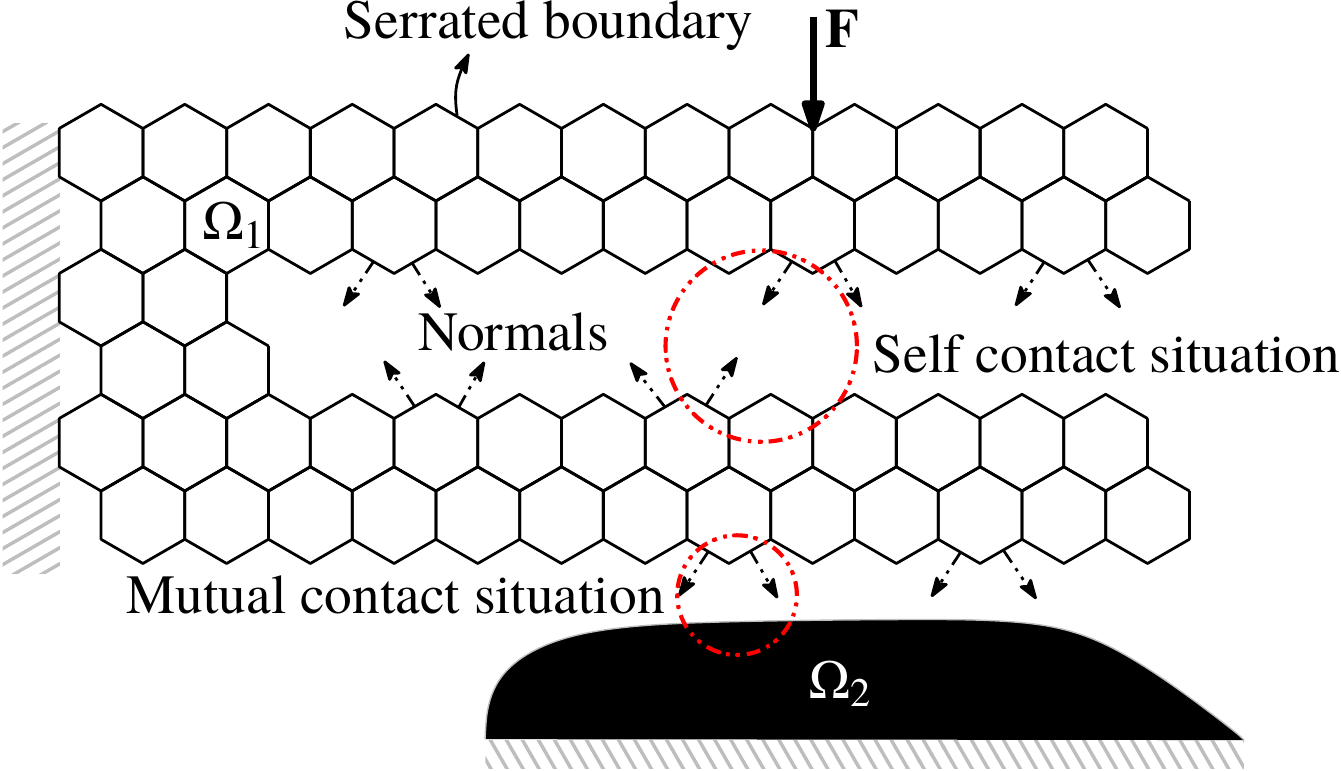}
		\caption{}
		\label{fig:Boundarysmoothing_a}
	\end{subfigure}
	\qquad
	\begin{subfigure}{0.45\textwidth}
		\centering
		\includegraphics[scale=0.5]{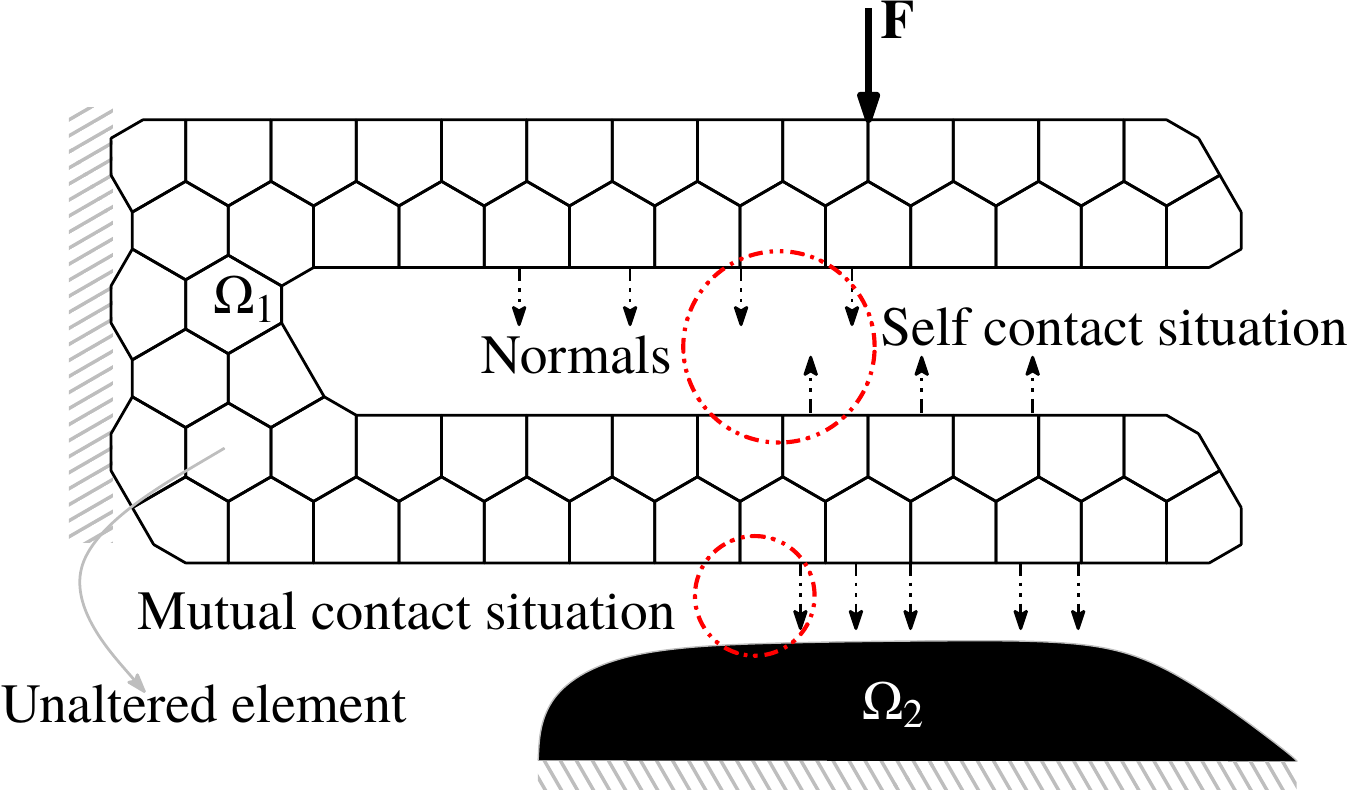}
		\caption{}
		\label{fig:Boundarysmoothing_b}
	\end{subfigure}
	\caption{Two bodies $\Omega_1$ and $\Omega_2$ come into contact. (\subref{fig:Boundarysmoothing_a}) $\Omega_1$ without boundary smoothing and (\subref{fig:Boundarysmoothing_b}) $\Omega_1$ with boundary smoothing. Jumps in boundary normals are subdued with a boundary smoothing scheme. Self and mutual contact sites are depicted with dash-dotted red circles.}
	\label{fig:Boundarysmoothing}
\end{figure}

Hexagonal elements are used to parameterize the design domain. These elements provide edge connectivity between any two contiguous elements \citep{saxena2007honeycomb,langelaar2007use,saxena2008material,talischi2009honeycomb,saxena2011topology,kumar2015topology,singh2020topology} and thus, alleviate checkerboard patterns or alternating filled and void elements, and point connections naturally. Negative circular masks are employed to remove material and also, to generate contact surfaces within some of them \citep{kumar2016synthesis}. In cases wherein only material removal (e.g. self contact) is to be performed, an $i^\mathrm{th}$ mask is defined via its center coordinates $(x_i,\,y_i)$ and radius $r_i$. Two more parameters $(s_i,\,f_i)$ are included within the definition of the mask, if an external (rigid) contact surface is also to be generated. Herein, $s_i$ and $f_i$ are binary and real fraction ($0<f_i<1$) variables, respectively. $s_i =1$ indicates generation of a contact surface with radius $f_ir_i$ within the mask, whereas $s_i = 0$ means that no contact surface is generated. The material state of each element toggles between void, $\rho(\Omega_\mathrm{H})=0$, and solid, $\rho(\Omega_\mathrm{H})=1$,  phases as positions and sizes of negative circular masks get updated. In each optimization iteration, all unexposed elements, i.e., elements with $\rho(\Omega_\mathrm{H})=1$ constitute the potential candidate design (Fig.~\ref{fig:schematicformulation}). These designs contain many V-notches on their bounding surfaces (Fig.~\ref{fig:Boundarysmoothing_a}). A boundary resolution and smoothing scheme \citep{kumar2015topology}, which shifts boundary nodes systematically is implemented, so that normals of the boundaries become well-defined (Fig.~\ref{fig:Boundarysmoothing_b}). Mean value shape functions \citep{hormann2006mean} are employed for nonlinear finite element analysis. To evaluate contact forces and corresponding stiffness matrices, the augmented Lagrange multiplier method in conjunction with segment-to-segment contact approach is implemented \cite{kumar2017_diss}. The Newton-Raphson method is used to solve nonlinear mechanical equilibrium equations.

Prior to the analysis, a set of shape morphing nodes (SMNs) are selected in the design region.  Elements containing those nodes are determined and termed shape morphing elements (SMEs).  SMEs must always be a part of the potential intermediate candidate design, i.e., SMEs constitute a solid non-design region. The mutated negative masks which overlay on SMEs are shifted systematically such that all SMEs remain in their solid material state. The design vector is updated accordingly. An objective based on Fourier shape descriptors \citep{zahn1972fourier} is conceptualized to evaluate the error between the desired and actual shapes. The actual shape is defined by the updated nodal positions of SMNs after completion of the Newton-Raphson iterations. The objective is minimized by the stochastic hill-climber method \citep{kumar2015synthesis,kumar2017implementation}.

\section {Problem formulation}\label{Sec:Problemformulation}
\begin{figure}[h!]
	\centering
	\begin{subfigure}[t]{0.45\textwidth}
		\includegraphics[scale=1.25]{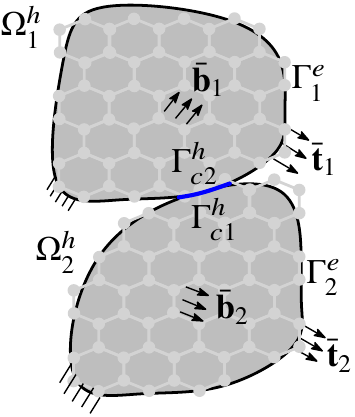}
		\caption{}
		\label{fig:contactfig2}
	\end{subfigure}
	\begin{subfigure}[t]{0.45\textwidth}
		\includegraphics[scale=0.850]{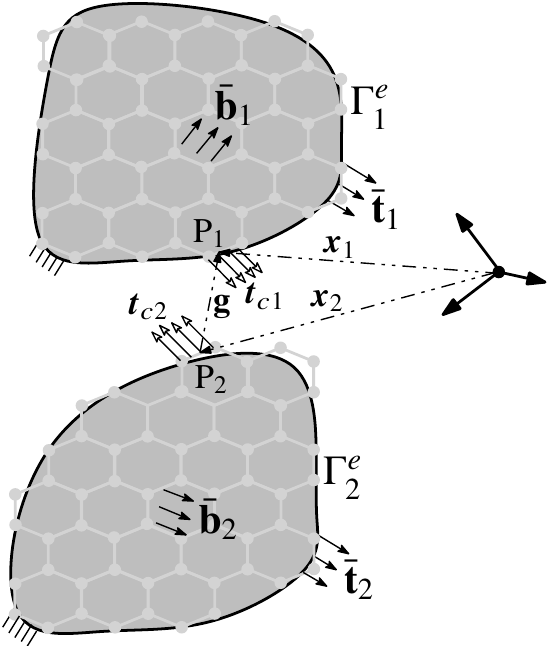}
		\caption{}
		\label{fig:contactfig3}
	\end{subfigure}
	\caption{Two bodies $\Omega_k|_{k=1,2}$ (in discrete setting, $\Omega_k^h|_{k=1,2}$) with known surface tractions $\bar{\mathbf{t}}_k$, volumetric body forces  $\bar{\mathbf{b}}_k$, and boundary conditions are depicted. When these bodies come into contact then contact surfaces $\Gamma_{ck}$ (or $\Gamma_{ck}^h$) and respective contact surface tractions $\bm{t}_{ck}$ appear. Consider points P$_1\in\Gamma_{c1}$ and P$_2\in\Gamma_{c2}$ with position vectors $\bm{x}_1$ and $\bm{x}_2$, respectively. Then the gap vector $\bm{g}$ is evaluated as $\bm{x}_2 - \bm{x}_1$.}
	\label{fig:contactfig}
\end{figure}

When two bodies interact, they experience contact forces at their respective contact boundary facets (Fig.~\ref{fig:contactfig3}). Typically, these forces depend on boundary normals \citep{wriggers2006computational}. Jumps in normals  are undesirable because they lead to non-convergence in contact analysis \citep{wriggers2006computational}. Serrated boundary facets lead to discontinuity in boundary normals (Fig.~\ref{fig:Boundarysmoothing_a}). To subdue serrations from the bounding surfaces, the boundary resolution and smoothing scheme \citep{kumarembedded,kumar2015topology} is incorporated.
\subsection {Boundary resolution and smoothing}\label{Sec:Boundary smoothing and resolution}
The boundary resolution and smoothing is accomplished in two steps.  In the first, identification of boundary edges, which are not shared by two or more elements, is performed and hence, boundary nodes constituting such edges are recognized. In the second, boundary nodes are projected along their shortest perpendiculars on the straight segments joining the mid-points of the boundary edges.  Such projections can be performed multiple times on the updated nodal coordinates \citep{kumar2015topology}.

Hexagonal elements are removed in two steps: (1) Elements which are overlaid by negative masks are removed and then, the smoothing is performed. (2) Elements which are not affected (Fig.~\ref{fig:Boundarysmoothing_b}) by smoothing in the former step are also removed subsequently. This is equivalent to placing additional negative masks over such elements. The latter step helps achieve candidate CMs with   slender members and thus, facilitate in their large deformation. Additionally,  overall volume of the mechanism is reduced. At the end of the second step, all  remaining hexagonal elements are considered in their regular shape, and boundary smoothing is performed again before contact finite element analysis is executed. As a consequence of the boundary smoothing, some boundary elements morph into concave elements \citep{kumar2015topology}. Mean value shape functions \citep{hormann2006mean} which can cater to generic polygonal shape finite elements are used for finite element analysis. For the sake of completeness, we briefly present the employed contact finite element formulation.
\subsection{Contact finite element formulation} \label{Sec:Contactformulation}

To evaluate contact forces and corresponding contact stiffness matrices, frictionless and adhesionless contact is assumed. Contact is modeled using the augmented Lagrange multiplier method in association with the Uzawa type \citep{bertsekas2014constrained} algorithm while considering the segment-to-segment approach. Classical penalty method is employed in the inner loop whereas in the outer loop, the Lagrange multiplier is updated \citep{wriggers2006computational}. In the classical penalty method, the contact traction $\bm{t}_c$ is defined as
\begin{equation} \label{Eq:contacttraction}
\bm{t}_{c} =
\begin{cases}
-\epsilon_n g_n\bm{n}_\text{p} \qquad \,\,\,\,\text{for}\,\,g_n<0 \\
\mathbf{0} ~\qquad  ~\qquad \quad \text{for}\,\,g_n\ge0
\end{cases}
\end{equation} 
where $g_n=(\bm{x} -\bm{x}_\text{p})\cdot\bm{n}_\text{p}$ is the normal gap. $\bm{x}_\text{p}$ is the projection point of $\bm{x}\in\Gamma_{c1}^h$ on the surface $\Gamma_{c2}^h$. $\bm{n}_{p}$ is the unit normal at the projection point $\bm{x}_\text{p}$ which is determined by solving the following minimization problem
\begin{equation}\label{Eq:Projectionpoint}
\bm{x}_\text{p} = \{\bm{x_2}: \min_{\bm{x}_2\in\Gamma_{c2}^h}||\bm{x}-\bm{x}_2||\,\, \forall \bm{x}\in \Gamma_{c1}^h\}.
\end{equation}

\noindent In a finite element setting, the virtual work contribution of elemental contact forces can be written as
\begin{equation} \label{Eq:Elementalcontactforce}
\mathbf{f}_{c}^e = - \int_{\Gamma_c^e} \mathbf{N}^\mathrm{T}\bm{t}_{c}^e\, \mathrm{d}a,
\end{equation}
and by assembling all such $\mathbf{f}_{c}^e$, one can find the global contact force $\mathbf{f}_{\text{c}}$. $\mathbf{N} = [N_1\bm{I},\, N_2\bm{I}]$ with $N_1 =\frac{1}{2}(1-\xi)$, $N_2 =\frac{1}{2}(1+\xi)$ and $\xi \in[-1,\,1]$. Further, $\text{d}a$ is the elemental area and $\bm{I}$ is the identity tensor.

The discretized weak form of the mechanical equilibrium equations then leads to the global finite element equilibrium equations 
\begin{equation} \label{Eq:mechanicalEquilibrium}
\mathbf{f}(\mathbf{u}) = \mathbf{f}_{\mathrm{int}}+ \mathbf{f}_{\mathrm{c}}-\mathbf{f}_{\mathrm{ext}}  = \mathbf{0},
\end{equation} 
where $\mathbf{f}_{\mathrm{int}},\,\mathbf{f}_{\mathrm{c}},\,\text{and}\,\mathbf{f}_{\mathrm{ext}}$ are the internal, contact and external forces, respectively. Eq.~\ref{Eq:mechanicalEquilibrium} is solved using the Newton-Raphson iterative method. One evaluates the elemental internal force $\mathbf{f}^e_\mathrm{int}$ as
\begin{equation} \label{Eq:elementalinternalforce}
\mathbf{f}^e_\mathrm{int} = \int_{\Omega_{k}^h}\mathbf{B}_\mathrm{UL}^\mathrm{T}\bm{\sigma}\, \mathrm{d}v,
\end{equation}
where $\mathbf{B}_\mathrm{UL}$ is the discrete strain-displacement matrix \citep{bathe2006finite} of an element in the current configuration\footnote{using the updated Lagrangian formulation}, and $\text{d}v$ is the elemental volume. For 2D cases such as plane strain, one can evaluate $\text{d}a = t\,\text{d}s$ and $\text{d}v = t\,\text{d}a$, where t is the thickness and $\text{d}s$ is the arc segment. $\bm{\sigma}$ is the Cauchy stress tensor evaluated using the nonlinear, isotropic, neo-Hookean material model \citep{zienkiewicz2005finite} 

\begin{align}\label{Eq:Cauchystress}
\bm{\sigma} &= \frac{\mu}{J}(\bm{F}\bm{F}^\mathrm{T}-\bm{I})+\frac{\lambda}{J}(\ln J)\bm{I},\qquad \  J = \text{det}\,\bm{F}
\end{align}
where $\mu = {E}/{2/(1+\nu)}$ and $\lambda = {2\mu\nu}/({1-2\nu})$ are Lame's constants, and $\bm{F} = \text{Grad}\, \bm{u} + \bm{I}$ is the deformation gradient. Further, $E$ and $\nu$ are Young's modulus and Possions' ratio, respectively. 

\subsection{Formulation of objective function and optimization problem}\label{Sec:Objectiveformulation}
An objective based on  Fourier Shape Descriptors (FSDs) \citep{zahn1972fourier} is formulated and minimized. This objective lets a user to exercise individual control on the errors in shape, size and initial orientation between two curves \citep{ullah1997optimal}. First, a  curve is closed in the clockwise sense such that it does not intersect itself. Then its Fourier coefficients are evaluated wherein the curve is parameterized using its normalized arc length. 

Let $A_n^k$ and $B_n^k$ be the Fourier coefficients, $\theta^k$ and $L^k$ be the initial orientation and total length of the two curves, $k= a\,,\, d\,$ represent the actual and desired shapes, respectively. Further, $n$ is the total number of Fourier coefficients. One evaluates the FSDs objective as
\begin{equation} \label{Eq:21}
f_0(\mathbf{v}) = \lambda_a A_\mathrm{{err}}+ \lambda_b B_\mathrm{{err}}+ \lambda_L L_\mathrm{{err}}+ \lambda_\theta \theta_\mathrm{{err}},
\end{equation}
where $\lambda_a,\,\lambda_b,\,\lambda_L,\,\text{and}\,\lambda_\theta$ are user defined weight parameters for the errors 
\begin{equation} \label{Eq:22}
\begin{aligned}
A_\mathrm{{err}}&=\sum_{i=1}^{n}(A_i^d-A_i^a)^2,\qquad
B_\mathrm{{err}}=\sum_{i=1}^{n}(B_i^d-B_i^a)^2,\,\\
L_\mathrm{{err}}&=(L^d-L^a)^2,\qquad \quad
\theta_\mathrm{{err}}=(\theta^d-\theta^a)^2,
\end{aligned}
\end{equation}
and $\mathbf{v}$ is the design vector. The units of the $\lambda$'s are chosen such that $f_0$ is dimensionless.The optimization problem then is
\begin{equation}\label{Eq:optimization}
\begin{aligned}
& \underset{\mathbf{v}}{\text{min}}
& &f(\mathbf{v}) + \lambda_v (V-V^*),\\
& \text{such that}, & & \mathbf{f}(\mathbf{u}) = \mathbf{0};\,\, q_L\le q_i\le q_U|_{q_i = x_i,\,y_i,\,r_i}\\ 
&\,& &s_i\,(= 0\,\, \text{or}\,\, 1)\,;\,\, f_i\,[\in (0,1)]
\end{aligned}
\end{equation}
where $V^*$ and $V^c$ are the desired  and current volumes of the CSMCM, and $\lambda_v$ is the volume penalization parameter. $\lambda_v = 0$ is taken, when $V^*<V^c$, otherwise $\lambda_v = 20$ is used. $q_L$ and $q_U$ denote the lower and upper limits for $q_i\in\mathbf{v}$.
\subsection{Hill climber search }\label{Sec:Hillclimbersearch}
Let the total number of overlaid negative circular masks  be N$_m$. Each mask is defined via its $x,\,y,\,r,\,s,\,\text{and}\,f$ variables. The design vector $\mathbf{v}$ consists of $5$N$_m$ variables. One sets a probability parameter $pr\,(=0.08)$ for each variable $d \in \mathbf{v}$. In each optimization iteration, one generates a random number $\chi$. If $\chi<pr$, the corresponding variable is altered as $d_\mathrm{new} = d_\mathrm{old} \pm (\kappa\times m)$, where $0<\kappa<1$ is a random number and $m$ is set to $10\%$ of the domain size, $\max(L_1,\,L_2)$. This mutation leads to a new design vector $\mathbf{v}_\mathrm{new}$. $s_i$ which indicates a contact surface generation is mutated as, if $\chi<pr$ and $\kappa<0.50$, $s_i= 1$, else $s_i=0$. Likewise, $f_i\in[0,\,1]$ is also mutated. The magnitude of input force $F$ is also taken as a design variable \citep{mankame2007synthesis} and updated as $F_\text{new} = F_\text{old}\pm (\kappa\times m)$. At this instance, if the input location, output location (member) and some fixed (boundary) conditions are available in the new design, then one evaluates the FSDs objective $f_\mathrm{new}$ as per design vector $\mathbf{v}_\mathrm{new}$, otherwise the design is penalized. If $f_\mathrm{new}<f_\mathrm{old}$, the design vector is updated. The process is continued until the maximum number of iterations is reached or terminated when it is found that the change in objective value for 10 successive optimization iteration is less than $\Delta f = 0.01$.
\begin{table}[h!]
	\centering
	\begin{tabular}{  l  c  c  }
		\hline
		\textbf{Parameter's name} & \textbf{Units} & \textbf{Value}  \\ \hline
		Design domain  &---& 30$\Omega_\mathrm{H}\times$30$\Omega_\mathrm{H}$  \\ 
		Maximum radius of $\Omega_\mathrm{M}$ & mm& 8.0\\
		Minimum radius of $\Omega_\mathrm{M}$ & mm& 0.1\\
		Maximum \# of iterations &---& 5000 \\ 
		Young's modulus ($E$) & MPa & 2100 \\ 
		Poisson's ratio &--- & 0.33 \\ 
		Permitted volume fraction ($\frac{V^*}{V}$)&---& 0.30\\
		Mutation probability &---& 0.08 \\ 
		Contact surface radii factor &---& 0.75 \\ 
		Maximum mutation size ($m_\mathrm{{max}}$) &---& 5\\ 
		Upper limit of the load ($\bm{F}_\mathrm{Upp}$) & N& 1000\\ 
		Lower limit of the load ($\bm{F}_\mathrm{Low}$) & N& -1000\\ 
		Weight of $a_\mathrm{err}$ ($\lambda_\mathrm{a}$) & \si{\per\radian\squared}& 100 \\ 
		Weight of $b_\mathrm{err}$ ($\lambda_\mathrm{b}$) & \si{\per\radian\squared}& 100\\ 
		Weight of  length error ($\lambda_\mathrm{L}$) & \si{\per\milli\meter}& 1\\ 
		Weight of orientation error ($\lambda_\mathrm{\theta}$) &\si{\per\radian\squared}& 1\\ 
		Boundary smoothing steps ($\beta$)& ---& 10 \\ 
		Penalty parameter ($\epsilon_n$)& N/mm$^{3}$& $60E/L_2$ \\
		Penalty parameter ($\epsilon_s$)& N/mm$^{3}$& $5E/L_2$\\ \hline
	\end{tabular} 
	\caption{Parameters used in the synthesis for Example~1, Example~2 and Example~3. $\epsilon_n$ and $\epsilon_s$ are the penalty parameters for mutual and self contact, respectively.}
	\label{T1}
\end{table}
\section{Numerical examples and discussion} \label{Sec:NumericalExamplesdiscusssion}

\begin{figure}[h!]
	\begin{subfigure}[t]{0.5\textwidth}
		\centering
		\includegraphics[scale=1]{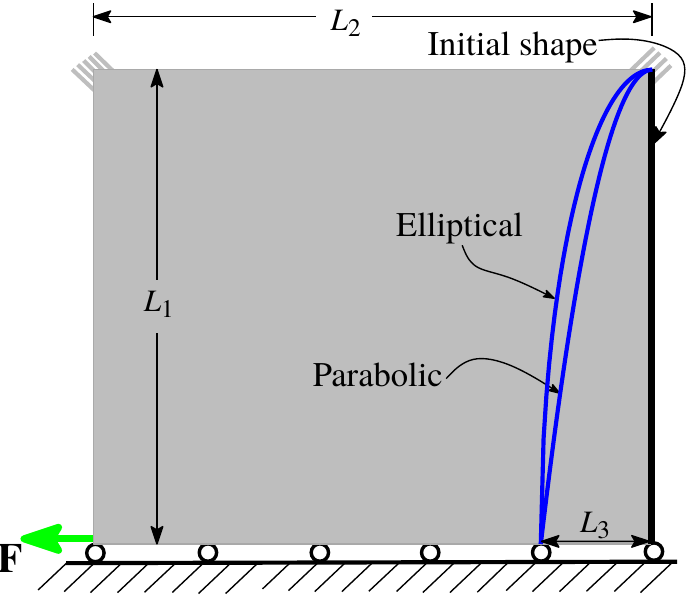}
		\caption{}
		\label{fig:prob1_2}
	\end{subfigure}
	\begin{subfigure}[t]{0.5\textwidth}
		\centering
		\includegraphics[scale=1]{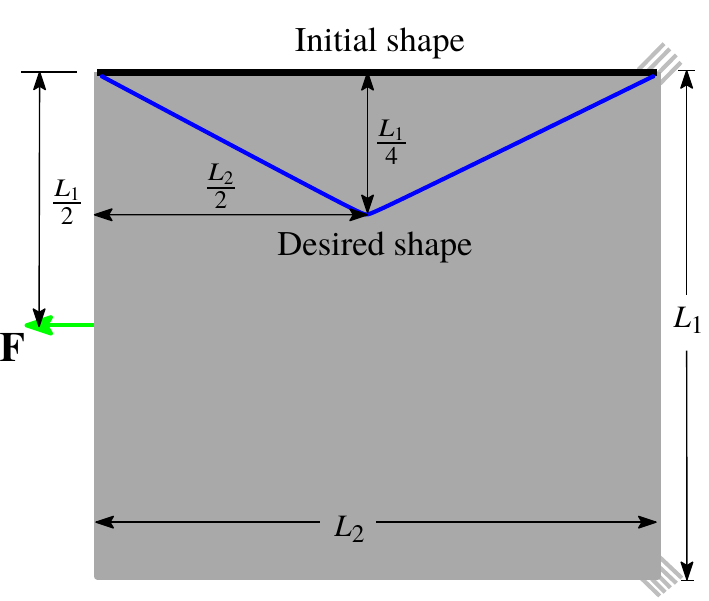}
		\caption{}
		\label{fig:prob3}
	\end{subfigure}
	\begin{subfigure}[t]{0.5\textwidth}
		\centering
		\includegraphics[scale=1]{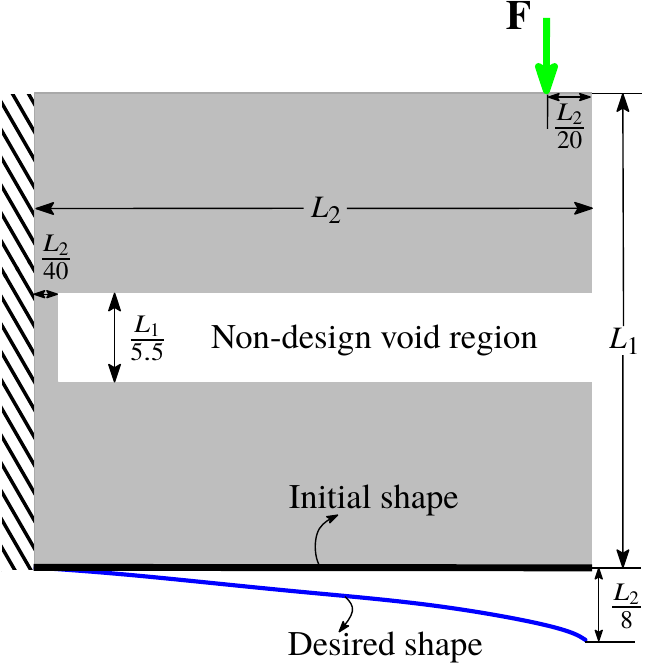}
		\caption{}
		\label{fig:prob4}
	\end{subfigure}
	\caption{ (\subref{fig:prob1_2}) Design specification for Example~1 and Example~2, (\subref{fig:prob3}) Design specification for Example~3 and (\subref{fig:prob4}) Design specification for Example~4. For Example~1, Example~2 and Example~3, $L_1 = 30\frac{3}{2}\si{\milli\meter},\, L_2 = 30\sqrt{3}\si{\milli\meter},\,L_3 = \frac{L_2}{3\sqrt{3}}\si{\milli\meter}$. For Example~4, $L_1 = 39\frac{3}{2}\si{\milli\meter},\, L_2 = 40\sqrt{3}\si{\milli\meter}$.}
	\label{fig:Probdesignspecificcation}
\end{figure}

\begin{figure}[h!]
	\centering
	\begin{subfigure}[t]{0.45\textwidth}
		\centering
		\includegraphics[scale=1.75]{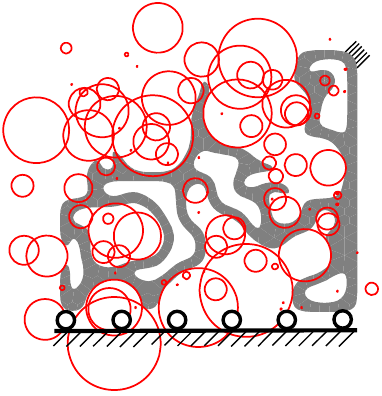}
		\caption{}
		\label{fig:Example1solution}
	\end{subfigure}
	\begin{subfigure}[t]{0.45\textwidth}
		\centering
		\includegraphics[scale=1.85]{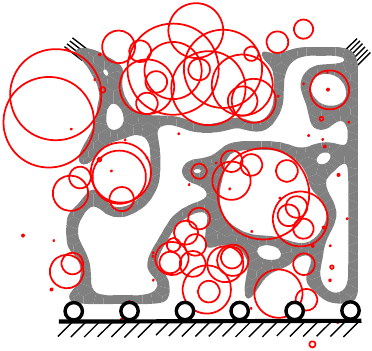}
		\caption{}
		\label{fig:Example2solution}
	\end{subfigure}
	\begin{subfigure}[t]{0.45\textwidth}
		\centering
		\includegraphics[scale=1.75]{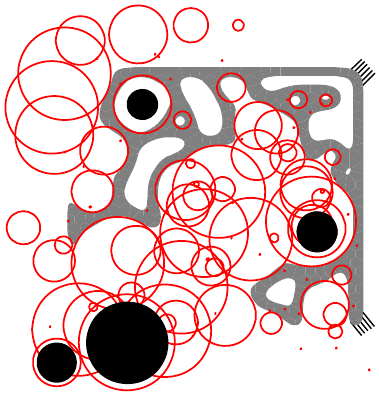}
		\caption{}
		\label{fig:Example3solution}
	\end{subfigure}
	\begin{subfigure}[t]{0.45\textwidth}
		\centering
		\includegraphics[scale=1]{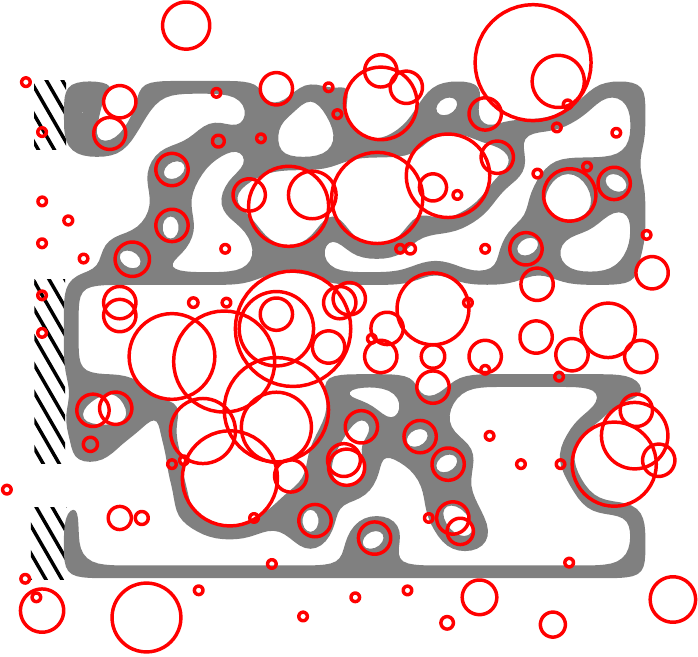}
		\caption{}
		\label{fig:Example4solution}
	\end{subfigure}
	\caption{Solutions to Example~1, Example~2, Example~3 and Example~4 with  boundary conditions are shown in figures (a), (b), (c) and (d), respectively. The final positions and sizes of circular masks (red) are also depicted.}	
	\label{fig:Solutions}
\end{figure}

Efficacy of the presented method is demonstrated via four contact-aided shape morphing  compliant mechanisms which are synthesized for different prescribed shapes (i.e. parabolic,  elliptical, and V-shape) shown in Fig.~\ref{fig:Probdesignspecificcation}. The design specification is also depicted and various parameters are tabulated in Table \ref{T1}. Plane-strain condition is assumed. The total number of Fourier coefficients is fixed to 50. The active set strategy in conjunction with contact-pairs detection scheme presented in \cite{kumar2019computational} is used to determine activeness and inactiveness of self and mutual contact modes. 

\subsection{Example 1}
The symmetric half design domain specifications and the desired parabolic profile for this example are depicted in Fig.~\ref{fig:prob1_2}. The left and right corners of the top edge of the symmetric half design domain are fixed (Fig.~\ref{fig:prob1_2}). To achieve the optimized compliant mechanism, $12$ masks in horizontal and $8$ masks in the vertical directions are employed. Only self contact is permitted and hence, masks are not required to generate rigid contact surfaces.  

\begin{figure}[h!] 
	\includegraphics[width=\textwidth]{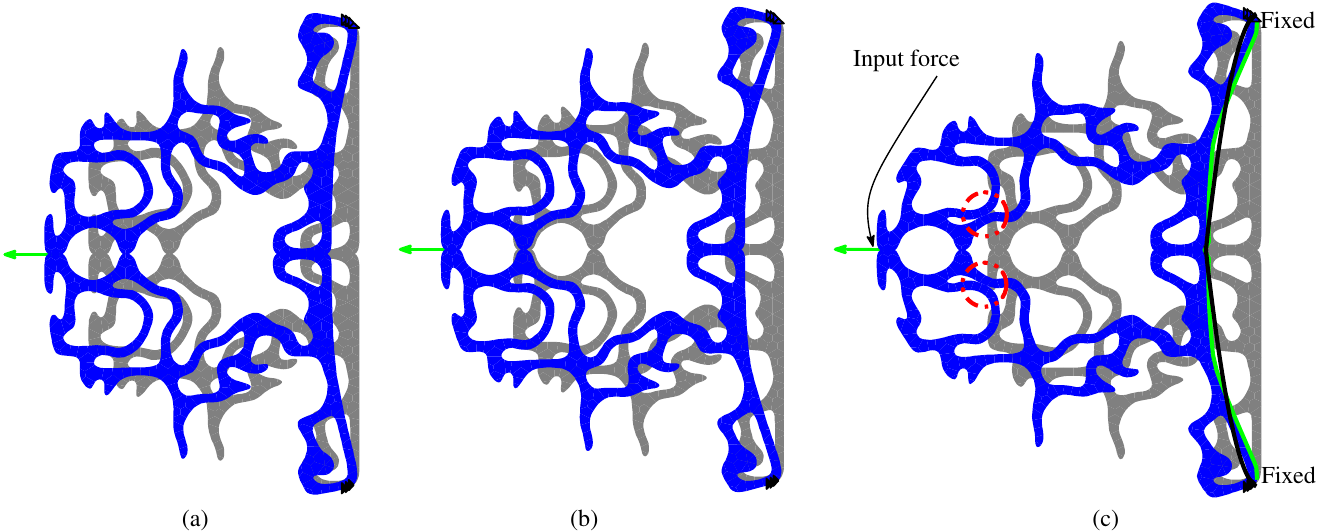}
	\caption{Example 1: Three deformed configurations (blue) are overlayed on the undeformed mechanism (gray). Figure (c) depicts the desired (black curve) and actual (green curve) shapes of the specified vertical member. Active contact locations are depicted using dash-dotted red circles. The input force and boundary conditions are also shown.}
	\label{fig:example1deformed}
\end{figure}

The final solution  is obtained after $715$ optimization iterations. The final symmetric half  result is suitably converted into a full mechanism (Fig.~\ref{fig:Example1solution}). Various configurations at different deformed states are shown in Fig.~\ref{fig:example1deformed}. The figure also shows two locations of self contact encircled in dash-dotted red circles. The obtained optimum actuation force is $-100.59$ N in the horizontal direction. Self contact occurs much later in the deformation history and does not influence the actual shape much. One can notice that the final mechanism has some extra appendages that mechanically may not be contributing significantly and thus, those may be removed in the post processing step.

\subsection{Example 2}
The design specifications, optimization parameters, and the number of masks used are the same as those for Example~1, however, the final desired shape sought is elliptical (Fig.~\ref{fig:prob1_2}). 
\begin{figure}[h!]
	\includegraphics[width=\textwidth]{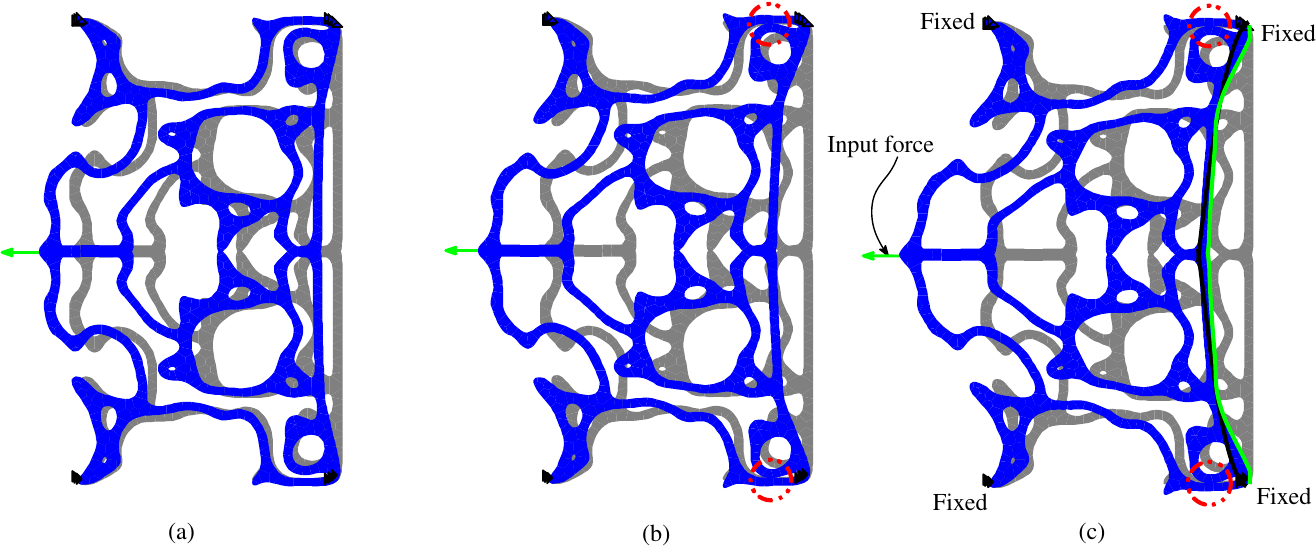}
	\caption{Example 2: Three deformed configurations (blue) are depicted with the undeformed mechanism (gray). The desired (black curve) and actual (green curve) shapes of the specified vertical member are shown in (c). Dash-dotted red circles are used to depict active contact locations. The input force and boundary conditions are also shown.}
	\label{fig:example2deformed}
\end{figure}
The optimized design is shown in Fig.~\ref{fig:Example2solution}. The final positions and shapes of the negative masks are also depicted. Deformed  configurations of the full mechanism at different states are shown in Fig.~\ref{fig:example2deformed}. While deforming, the mechanism experiences self contact at two locations (Fig.~\ref{fig:example2deformed}). The final mechanism is obtained after $782$ optimization iterations with $-96.64$ N actuating force in the horizontal direction. Self contact happens much earlier in the deformation history, which  helps achieve the actual elliptical profile, very close to the desired shape (Table~\ref{Table:FSDscomparasion}).

\subsection{Example 3}
The design specifications for the third example are shown in Fig.~\ref{fig:prob3}. The same design parameters are used as for Example 1.  The top and bottom corners of the right edge of the design domain are fixed. We take 10 masks in each direction for optimization. The masks are permitted to generate contact surfaces, i.e., 5 design parameters are used for each mask. This example is solved to achieve a V-shape for the specified edge (Fig.~\ref{fig:prob3}). Note  that in a continuum setting, getting such desired shapes is only possible if one uses very fine mesh and/or if there is discontinuity/notch at
that boundary. For coarse meshes this is an extreme test case for the proposed mechanism design methodology.
\begin{figure}[h!]
	\includegraphics[width=\textwidth]{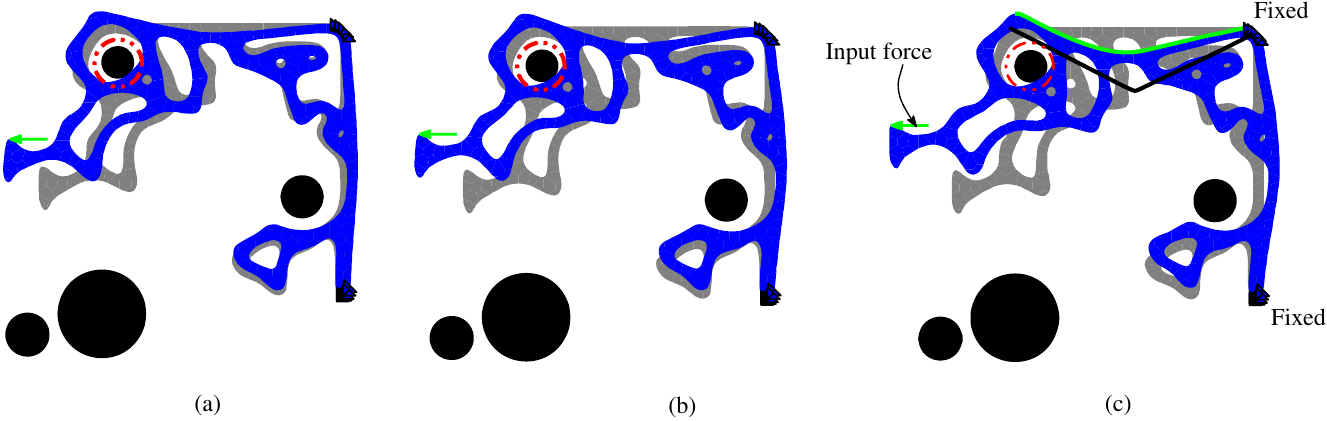}
	\caption{Example 3: Three deformed configurations (blue) along with input force and boundary conditions are overlayed on the undeformed mechanism (gray). Figure (c) depicts the desired (black curve) and actual (green curve) shapes of the specified horizontal member. The active contact surface is depicted with dash-dotted red circles.}
	\label{fig:example3deformed}
\end{figure}
The optimum solution (Fig.~\ref{fig:Example3solution}) is obtained after $947$ optimization iterations. The mechanism interacts with only one contact surface though many such surfaces are present (Fig.~\ref{fig:example3deformed}). The final input force in the horizontal direction is $-96.30$ N. Various deformed configurations with active contact locations, actuation force, boundary condition are depicted in Fig.~\ref{fig:example3deformed}. Herein, mutual contact occurs much earlier in the deformation history. It is reckoned that the relative frictionless slip between the rigid surface and the loop (top left) contributes significantly to achieving a shape close to the \textquoteleft V' profile. However, the desired \textquoteleft kink' is not observed, this is because continuum surface deformations are usually smooth despite the presence of contact. 
\subsection{Example~4}
The design domain specifications for Example~4 are displayed in Fig.~\ref{fig:prob4}. The left edge of the domain is fixed. A non-design void region of size $\frac{39L_2}{40}\times\frac{L_1}{5.5}$, symmetric to the center of the domain, as illustrated in the figure, is considered. An input force is applied on the top edge as depicted in Fig.~\ref{fig:prob4}. The initial and desired shapes of a potential contact-aided compliant mechanism are indicated in Fig.~\ref{fig:prob4}. $40\times39$ hexagonal elements are used to discretize the design domain. We employ 12 and 10 negative masks along the $x-$ and $y-$directions, respectively. Other design parameters are the same as those mentioned in Table~\ref{T1}. Masks are  permitted to remove only material but not to generate contact surfaces, thus contact can only occur as self contact between flexible members.
\begin{figure}[h!] 
	\includegraphics[width=\textwidth]{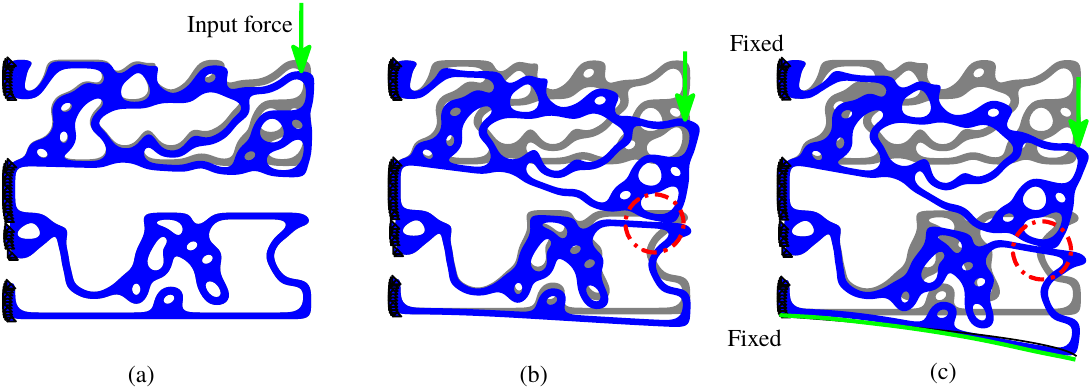}
	\caption{Three deformed configurations (blue) are overlayed on the undeformed mechanism (gray). Figure (c) depicts the desired (black curve) and actual (green curve) shapes of the specified horizontal member. Active contact locations are depicted using dash-dotted red circles. The input force and boundary conditions are also shown. The desired shape cannot be obtained unless there is contact between the top and bottom subregions.}
	\label{fig:example4deformed}
\end{figure}
The final solution of the mechanism with final shape and size of the masks with their position is shown in Fig.~\ref{fig:Example4solution}. Different configurations at different instances are illustrated in Fig.~\ref{fig:example4deformed}. Self contact sites are also depicted in the figure using the dash-dotted red circles. The final input force is $-$\SI{25.0}{\newton} in the $y-$direction. One can notice that the mechanism achieves the desired profile while deforming due to self contact. Thus, self contact can play a vital role in achieving the final desired profile of a SMCM.

\subsection{Comparison between the desired and actual curves}
The error in shape and size between the two curves is formulated with respect to respective Fourier coefficients in terms of $R_m = \sqrt{A_m^2+ B_m^2}$, where $R_m|_{(m = 1,2,\cdots, n)}$ are curve invariants~\citep{zahn1972fourier}.  The overall relative change in shape $\zeta_{s}$ is evaluated as 
\begin{equation}
\zeta_s = \bigg[\frac{1}{n} \sum_{m=1}^{n} \frac{|R_m^d - R_m^a|}{R_n^d}\bigg],
\end{equation} 
where $R_m^d$ and $R_m^a$ are  invariants corresponding to the desired and actual curves respectively. Likewise, the relative change in lengths is evaluated as $\zeta_l = \frac{|L^d - L^a|}{L^d}$. 

\begin{table}[h!]
	\centering
	\caption{Percentage change in FSD coefficients and length of actual curve of CSMCMs with respect to their corresponding desired curves} 
	\label{Table:FSDscomparasion}
	\begin{tabular}{  l    c  c } 
		\hline
		\textbf{Mechanisms} & \textbf{$\zeta_s$} (\%) & \textbf{$\zeta_l$}  (\%)  \\ \hline
		Example 1  & 0.394 & 4.645\\ 
		Example 2 & 0.233 & 5.962\\ 
		Example 3  & 0.557 & 12.722\\
		Example 4  & 1.99 & 3.12\\
		\hline
	\end{tabular} 
\end{table}  

Table~\ref{Table:FSDscomparasion} depicts the comparison of $\zeta_{s}$ and $\zeta_{l}$ for the presented examples. One notices for each problem $\zeta_s $ is within 2\% (Table~\ref{Table:FSDscomparasion}), indicating good shape agreement between the actual and desired curves. We notice 12.72\% length error between the desired and actual curves for Example~3.

\subsection{Verification of the deformed profiles}
ABAQUS is used to appraise the accuracy of the presented design approach by comparing the deformed profiles for the optimized designs with those obtained by ABAQUS analyses. 

\begin{table}[h!]
	\centering
	\caption{Percentage change in FSDs coefficients and length of actual curve of CSMCMs with respect to their corresponding curves obtained using ABAQUS} 
	\label{Table:FSDsABAQUS}
	\begin{tabular}{  l    c  c } 
		\hline
		\textbf{Mechanisms} & \textbf{$\zeta_s$} (\%) & \textbf{$\zeta_l$}  (\%)  \\ \hline
		Example 1  & 0.1808 & 7.043\\ 
		Example 2 & 0.1126 & 5.817\\ 
		Example 3  & 0.047 & 0.67\\
		Example 4  & 1.97 & 3.01\\
		\hline
	\end{tabular} 
\end{table} 

\begin{figure}[h!]
	\begin{subfigure}[t]{0.45\textwidth}
		\centering
		\begin{tikzpicture} 	
		\pgfplotsset{compat = 1.3}
		\begin{axis}[
		width = 1\textwidth,
		xlabel=\si{[mm]},
		ylabel= \si{[mm]},
		legend style={at={(0.5,0.5)},anchor=west}]
		\pgfplotstableread{R1actual.txt}\mydata;
		\addplot[smooth,mark=*,blue,mark size=1pt]
		table {\mydata};
		\addlegendentry{Actual}
		\pgfplotstableread{R1abaqus.txt}\mydata;
		\addplot[smooth,mark=square*,black,mark size=1pt]
		table {\mydata};
		\addlegendentry{ABAQUS}
		\end{axis}
		\end{tikzpicture}
		\caption{}
		\label{fig:result1comparision}
	\end{subfigure}
	\begin{subfigure}[t]{0.45\textwidth}
		\centering
		\begin{tikzpicture} 	
		\pgfplotsset{compat = 1.3}
		\begin{axis}[
		width = 1\textwidth,
		xlabel=\si{[mm]},
		ylabel= \si{[mm]},
		legend style={at={(0.5,0.5)},anchor=west}]
		\pgfplotstableread{R2actual.txt}\mydata;
		\addplot[smooth,mark=*,blue,mark size=1pt]
		table {\mydata};
		\addlegendentry{Actual}
		\pgfplotstableread{R2abaqus.txt}\mydata;
		\addplot[smooth,mark=square*,black,mark size=1pt]
		table {\mydata};
		\addlegendentry{ABAQUS}
		\end{axis}
		\end{tikzpicture}
		\caption{}
		\label{fig:result2comparision}
	\end{subfigure}
	\begin{subfigure}[t]{0.45\textwidth}
		\centering
		\begin{tikzpicture} 	
		\pgfplotsset{compat = 1.3}
		\begin{axis}[
		width = 1\textwidth,
		xlabel=\si{[mm]},
		ylabel= \si{[mm]},
		legend pos=north east]
		\pgfplotstableread{R3actual.txt}\mydata;
		\addplot[smooth,mark=*,blue,mark size=1pt]
		table {\mydata};
		\addlegendentry{Actual}
		\pgfplotstableread{R3abaqus.txt}\mydata;
		\addplot[smooth,mark=square*,black,mark size=1pt]
		table {\mydata};
		\addlegendentry{ABAQUS}
		\end{axis}
		\end{tikzpicture}
		\caption{}
		\label{fig:result3comparision}
	\end{subfigure}
	\begin{subfigure}[t]{0.45\textwidth}
		\centering
		\begin{tikzpicture} 	
		\pgfplotsset{compat = 1.3}
		\begin{axis}[
		width = 1\textwidth,
		xlabel=\si{[mm]},
		ylabel= \si{[mm]},
		legend pos=north east]
		\pgfplotstableread{R4actual.txt}\mydata;
		\addplot[smooth,mark=*,blue,mark size=1pt]
		table {\mydata};
		\addlegendentry{Actual}
		\pgfplotstableread{R4abaqus.txt}\mydata;
		\addplot[smooth,mark=square*,black,mark size=1pt]
		table {\mydata};
		\addlegendentry{ABAQUS}
		\end{axis}
		\end{tikzpicture}
		\caption{}
		\label{fig:result4comparision}
	\end{subfigure}
	\caption{The deformed profiles of the actual curves for Example 1, Example 2, Example 3 and Example~4 and those obtained using ABAQUS are depicted  in (\subref{fig:result1comparision}), (\subref{fig:result2comparision}), (\subref{fig:result3comparision}), and  (\subref{fig:result4comparision}), respectively. Horizontal and vertical axes of the plots represent  $x-$ and $y-$positions of the curves, respectively.}
	\label{fig:abaqusactual}
\end{figure}
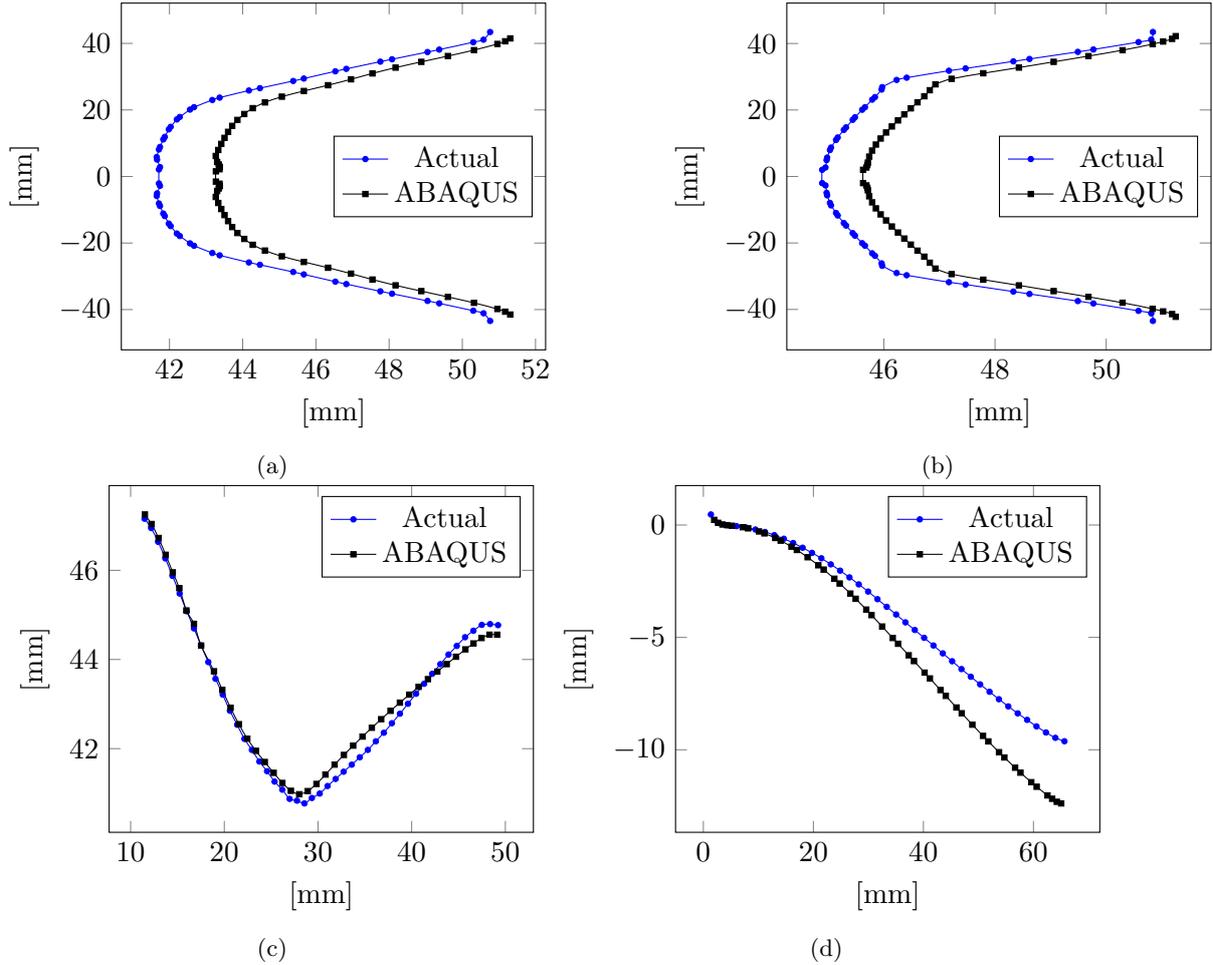

To perform the ABAQUS nonlinear contact analyses, (i) the optimal forces, (ii) boundary conditions, and (iii) active contact locations (self and/or mutual) of the optimized solutions (Fig.~\ref{fig:Solutions}) in association with the neo-Hookean material model, are used. Using the information of boundary nodes, the optimized results are converted into respective CAD models.  Four-noded plain-strain elements (CPE4I) are employed to describe the extracted CAD model of the mechanism. The actual profiles and those obtained using ABAQUS for the respective examples, are depicted in  Fig.~\ref{fig:abaqusactual}. The analyses indicate that the obtained deformed shapes closely follow the respective actual deformed shapes for the presented examples (Fig.~\ref{fig:abaqusactual} and Table~\ref{Table:FSDsABAQUS}).

\subsection{Influence of friction}
In this section, we present a study of frictional contact surfaces on the performance (the ability to obtain the desired deformed profiles) of the final mechanisms in ABAQUS by  considering different friction coefficients. The presented topology optimization approach though does not account for frictional contact, it can be readily added using the formulation mentioned in \citep{sauer2015unbiased}.  

\begin{table}[h!]
	\centering
	\caption{Percentage change in FSD coefficients and length of curve obtained with friction to those without friction using ABAQUS}
	\label{Table:Frictionstudy}
	\begin{tabular}{c| c c| c c}
		\hline
		\multicolumn{1}{c|}
		{\textbf{Mechanisms}} & \multicolumn{2}{c|}
		{\textbf{$\mu_f =0.25$}} & \multicolumn{2}{c} {\textbf{$\mu_f =0.35$}} \\
		\textbf{}  & \textbf{$\zeta_s$} (\%) & \textbf{$\zeta_l$}  (\%) & \textbf{$\zeta_s$} (\%) & \textbf{$\zeta_l$}  (\%)\\ \hline
		Example 1  & 0.00023 & 0.0015 & 0.00030 & 0.0018\\
		Example 2  & 0.0003& 0.0016 & 0.00035 & 0.0017\\
		Example 3  & 0.0173 & 2.038 & 0.0207 & 2.1314\\
		Example 4  & 0.0054 & 0.083 & 0.0077 &0.12\\
		\hline
	\end{tabular}
\end{table}

\begin{figure}[h!]
	\begin{subfigure}[t]{0.45\textwidth}
		\centering
		\begin{tikzpicture} 	
		\pgfplotsset{compat = 1.3}
		\begin{axis}[
		width = 1\textwidth,
		xlabel=\si{[mm]},
		ylabel= \si{[mm]},
		legend style={at={(0.5,0.5)},anchor=west}]
		\pgfplotstableread{R1abaqusf0.txt}\mydata;
		\addplot[solid, blue,line width=1pt]
		table {\mydata};
		\addlegendentry{$\mu_f=0.0$}
		\pgfplotstableread{R1abaqusf25.txt}\mydata;
		\addplot[loosely dashed,red,,line width=1pt]
		table {\mydata};
		\addlegendentry{$\mu_f=0.25$}
		\pgfplotstableread{R1abaqusf35.txt}\mydata;
		\addplot[densely dashdotdotted,black,,line width=1pt]
		table {\mydata};
		\addlegendentry{$\mu_f=0.35$}
		\end{axis}
		\end{tikzpicture}
		\caption{}
		\label{fig:result1friction}
	\end{subfigure}
	\begin{subfigure}[t]{0.45\textwidth}
		\centering
		\begin{tikzpicture} 	
		\pgfplotsset{compat = 1.3}
		\begin{axis}[
		width = 1\textwidth,
		xlabel=\si{[mm]},
		ylabel= \si{[mm]},
		legend style={at={(0.5,0.5)},anchor=west}]
		\pgfplotstableread{R2abaqusf0.txt}\mydata;
		\addplot[solid, blue,line width=1pt]
		table {\mydata};
		\addlegendentry{$\mu_f=0.0$}
		\pgfplotstableread{R2abaqusf25.txt}\mydata;
		\addplot[loosely dashed,red,,line width=1pt]
		table {\mydata};
		\addlegendentry{$\mu_f=0.25$}
		\pgfplotstableread{R2abaqusf35.txt}\mydata;
		\addplot[densely dashdotdotted,black,,line width=1pt]
		table {\mydata};
		\addlegendentry{$\mu_f=0.35$}
		\end{axis}
		\end{tikzpicture}
		\caption{}
		\label{fig:result2friction}
	\end{subfigure}
	\begin{subfigure}[t]{0.45\textwidth}
		\centering
		\begin{tikzpicture} 	
		\pgfplotsset{compat = 1.3}
		\begin{axis}[
		width = 1\textwidth,
		xlabel=\si{[mm]},
		ylabel= \si{[mm]},
		legend style={at={(0.5,1)},anchor=north west}]
		\pgfplotstableread{R3abaqusf0.txt}\mydata;
		\addplot[solid, blue,line width=1pt]
		table {\mydata};
		\addlegendentry{$\mu_f=0.0$}
		\pgfplotstableread{R3abaqusf25.txt}\mydata;
		\addplot[loosely dashed,red,,line width=1pt]
		table {\mydata};
		\addlegendentry{$\mu_f=0.25$}
		\pgfplotstableread{R3abaqusf35.txt}\mydata;
		\addplot[densely dashdotdotted,black,,line width=1pt]
		table {\mydata};
		\addlegendentry{$\mu_f=0.35$}
		\end{axis}
		\end{tikzpicture}
		\caption{}
		\label{fig:result3friction}
	\end{subfigure}
	\begin{subfigure}[t]{0.45\textwidth}
		\centering
		\begin{tikzpicture} 	
		\pgfplotsset{compat = 1.3}
		\begin{axis}[
		width = 1\textwidth,
		xlabel=\si{[mm]},
		ylabel= \si{[mm]},
		legend style={at={(0.5,1)},anchor=north west}]
		\pgfplotstableread{R4abaqusf0.txt}\mydata;
		\addplot[solid, blue,line width=1pt]
		table {\mydata};
		\addlegendentry{$\mu_f=0.0$}
		\pgfplotstableread{R4abaqusf25.txt}\mydata;
		\addplot[loosely dashed,red,,line width=1pt]
		table {\mydata};
		\addlegendentry{$\mu_f=0.25$}
		\pgfplotstableread{R4abaqusf35.txt}\mydata;
		\addplot[densely dashdotdotted,black,,line width=1pt]
		table {\mydata};
		\addlegendentry{$\mu_f=0.35$}
		\end{axis}
		\end{tikzpicture}
		\caption{}
		\label{fig:result4friction}
	\end{subfigure}
	\caption{The obtained deformed profiles of the actual curves with different frictional coefficients $\mu_f$ using ABAQUS are overlaid and depicted for Example~1, Example~2, Example~3 and Example~4 in (\subref{fig:result1friction}), (\subref{fig:result2friction}),  (\subref{fig:result3friction}), and (\subref{fig:result4friction}), respectively. Horizontal and vertical axes of the plots represent  $x-$ and $y-$positions of the curves, respectively.}
	\label{fig:abaqusfrictioncomparision}
\end{figure}
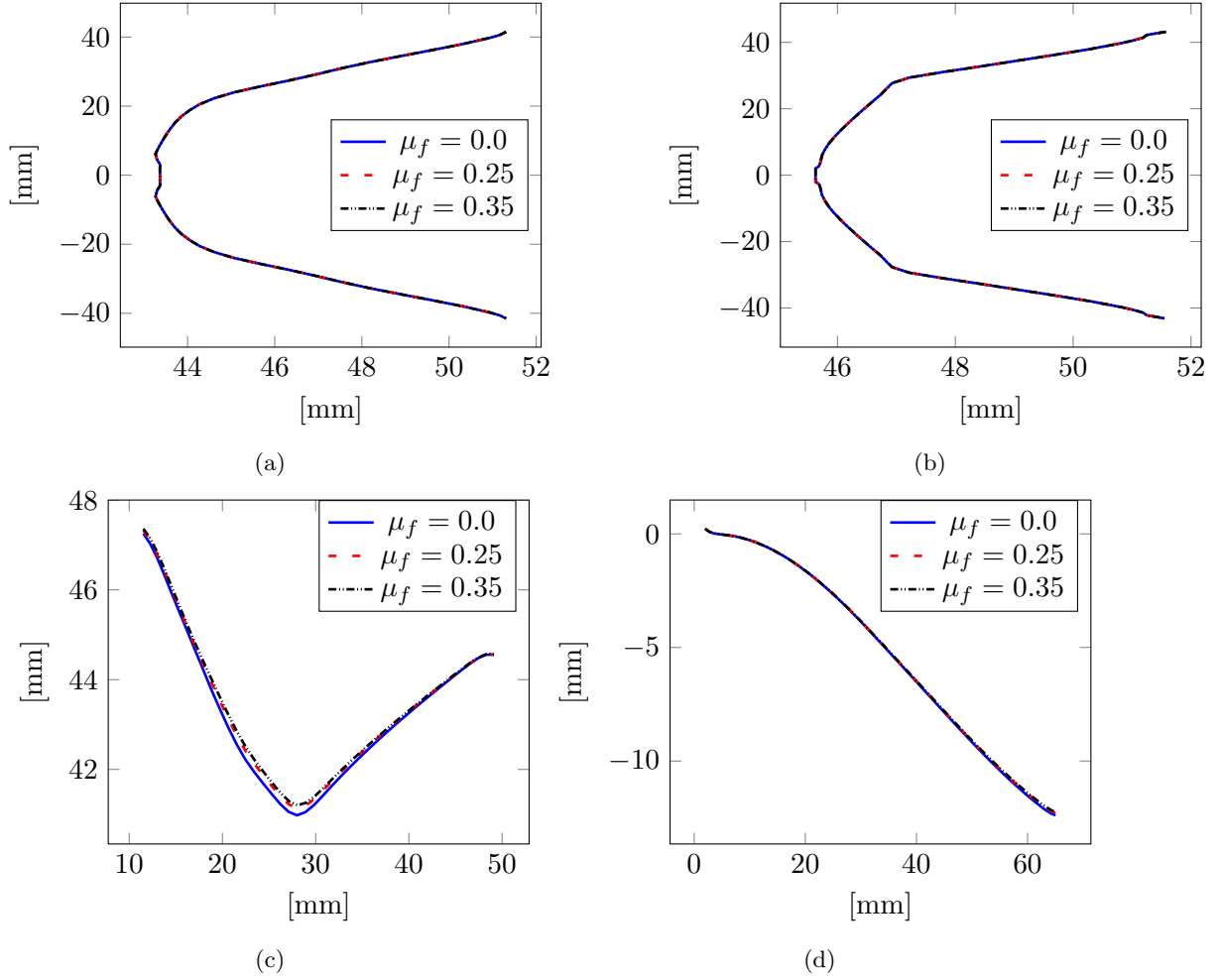
The deformed shapes of the pre-sepcified constituting members of the respective examples with  $\mu_f = 0$,\,$\mu_f = 0.25$ and $ \mu_f = 0.35$ are overlaid and compared in Fig.~\ref{fig:abaqusfrictioncomparision}. Percentage change in the FSDs coefficients and lengths of the deformed profiles with respect  to $\mu_f = 0$, are given in Table~\ref{Table:Frictionstudy}. One notices that friction does not alter the quantitative and/or qualitative behavior of the deformed shapes (Fig.~\ref{fig:abaqusfrictioncomparision} and Table~\ref{Table:Frictionstudy}). However, this may not be the case in a situation where contact surfaces are comparatively bigger in shape. 

\subsection{Presence of small holes}

One notices a few small holes in the final designs of Example~2, Example~3 and Example~4 (Fig.~\ref{fig:Solutions}). These holes appear due to the localized removal of either one/two FEs by (a group of) circular masks or  one/two unaffected FEs (Fig. \ref{fig:Boundarysmoothing_b}). The respective modified designs are obtained after removing those small holes. The nonlinear contact FE analyses in ABAQUS are performed using the final and their respective modified designs. The obtained end-compliance\footnote{End-compliance, $C$, is the compliance of a design in its equilibrium configuration. Mathematically, $C = \mathbf{f_\text{ext}^\text{T}\mathbf{u}}$, where $\mathbf{u}$ is the displacement vector of the design at the equilibrium position.} values are reported in Table~\ref{Table:compliance}. The differences in values of the end-compliance for all these examples are found to be below 1\%, thus those small holes can be removed in a post processing step.

\begin{table}[h!]
	\centering
	\caption{End-compliance values} \label{Table:compliance}
	\begin{tabular}{ l |c |c |c}
		\hline
		\multirow{2}{*}{\textbf{Mechanism}} & \textbf{Original design} & \textbf{Modified design} & \multirow{2}{*}{\textbf{Difference (\%)}} \\ \cline{2-3} 
		& \multicolumn{2}{c|}{End-compliance (Nmm)}        &                     \\ \hline
		Example 2                           &    1107.50                      &       1194.47                &            0.79         \\ \hline
		Example 3                           &      1093.97                    &     1088.20                  &          0.053           \\ \hline
		Example 4                           &       497.87                   &     495.66                  &               0.045      \\ \hline
	\end{tabular}
\end{table}

\section{Closure} \label{Sec:Closure}
An approach to synthesize  contact-aided shape morphing compliant mechanisms using hexagonal elements and negative circular masks, is presented. Self and/or mutual contact modes are permitted. Geometric and material nonlinearities are considered wherein a neo-Hookean material model is employed. Versatility of the presented method is demonstrated via four examples with various desired shapes. The optimized mechanisms for Example 1, Example 2 and Example 4 experience self contact while achieving their desired shapes, whereas mutual contact helps achieve the actual shape similar to the its desired one for Example 3. By and large, there is a good agreement between the desired and actual curves as differences in shape and size measure for these curves are within $1 \%$.

The augmented Lagrange multiplier method is used considering a segment-to-segment contact model. The implemented boundary smoothing reduces jumps in the normals of the boundary facets thereof and facilitates convergence of the contact analysis. The nonlinear mechanical equilibrium equations are solved using the Newton-Raphson method. An FSDs based objective is formulated and minimized, which permits to have individual control over the characteristics of a curve. Hill-climber, a zero-order search algorithm, is used. 

The optimized mechanisms are analyzed in ABAQUS using the  respective actuating force, boundary conditions, and active contact locations. It is noticed that the deformed profiles obtained by the approach and those by  ABAQUS are very close to each-other. Analyses considering frictional contact surfaces are also performed in ABAQUS. It is noted that friction does not alter the behavior of the deformed curves much. In future, we aim to design special characteristic mechanisms, e.g., with negative stiffness or zero-stiffness and statically balanced mechanisms in association with contact constraints, which can find applications in medical devices. 
	
	\bibliography{myreference}
	\bibliographystyle{spbasic} 
\end{document}